\documentclass[12pt, journal]{IEEEtran}
\usepackage{multirow}

\usepackage{cite}
\usepackage{flushend}

\usepackage[pdftex]{graphicx}
\ifCLASSINFOpdf
\else
\fi
\usepackage{amsmath, bm}
\usepackage{algorithmic}
\usepackage{algorithm}
\ifCLASSOPTIONcompsoc
  \usepackage[caption=false,font=normalsize,labelfont=sf,textfont=sf]{subfig}
\else
  \usepackage[caption=false,font=footnotesize]{subfig}
\fi
\usepackage{url}
\usepackage{setspace}
\onecolumn

\begin{document}
% Change title
\title{Representation Learning via \\Cauchy Convolutional Sparse Coding}

\author{Perla~Mayo,~%~\IEEEmembership{Member,~IEEE,}
        Oktay~Karakuş,~%\IEEEmembership{Member,~IEEE,}
        Robin~Holmes,~%~\IEEEmembership{Fellow,~OSA,}
        and~Alin~Achim,~%\IEEEmembership{Senior~Member,~IEEE}% <-this % stops a space
\thanks{P. Mayo, O. Karakuş and A. Achim are with the Visual Information Laboratory, University of Bristol, Bristol, UK, email: pm15334@bristol.ac.uk, o.karakus@bristol.ac.uk and Alin.Achim@bristol.ac.uk respectively.}% <-this % stops a space
\thanks{This work was supported in part by a CONACyT PhD studentship under grant 461322 (to Mayo), in part by the Engineering and Physical Sciences Research Council (EPSRC) under grant EP/R009260/1 (AssenSAR), and in part by a Leverhulme Trust Research Fellowship (to Achim).}
\thanks{Robin Holmes is with the NHS Trust Foundation.}% <-this % stops a space
}
\maketitle

% =======================================================
%               A B S T R A C T
\begin{abstract}
In representation learning, Convolutional Sparse Coding (CSC) enables unsupervised learning of features by jointly optimising both an \(\ell_2\)-norm fidelity term and a sparsity enforcing penalty. This work investigates using  a regularisation term derived from an assumed Cauchy prior for the coefficients of the feature maps of a CSC generative model. The sparsity penalty term resulting from this prior is solved via its proximal operator, which is then applied iteratively, element-wise, on the coefficients of the feature maps to optimise the CSC cost function.
%Such penalty term is solved by the proximal operator of the Cauchy sparsity function in an iterative shrinkage algorithm. 
The performance of the proposed Iterative Cauchy Thresholding (ICT) algorithm in reconstructing natural images is compared against the common choice of \(\ell_1\)-norm optimised via soft and hard thresholding.
%In this work a new approach aiming to learn features from data is proposed. The main pipeline followed is the one of Convolutional Sparse Coding in conjunction with an iterative shrinkage algorithm that we named Iterative Cauchy Thresholding (ICT). The algorithm works under the assumption that the coefficients describing the contribution of the features follow the Cauchy --a heavy tailed-- distribution with location 0, which is a valid assumption when seeking for sparsity.
ICT outperforms IHT and IST in most of these reconstruction experiments across various datasets, with an average PSNR of up to 11.30 and 7.04 above ISTA and IHT respectively.%, showing that, event though sparsity is not achieved in a similar fashion as done by the common thresholding techniques, the power in the learning of features is greater when the Cauchy proximal operator is used instead.
\end{abstract}

% =======================================================
%               K E Y W O R D S

% \fi
\IEEEpeerreviewmaketitle

% =======================================================
%               I N T R O D U C T I O N
% =======================================================
\section{Introduction}
%\IEEEPARstart{I}{n} 
Representation learning seeks to understand the underlying patterns and structures that give raise to the data of interest. This often involves using generative models to describe the processes involved in the formation of this data, using known or assumed priors \cite{bengio2013representation}. In some of these models it is assumed the data arises from of a linear operation among the elements of a set of basic or canonical features. Thus, in addition to selecting this set of features, it is also necessary to obtain their respective coefficients for generating any given sample. Computing the coefficients for such a sample effectively transforms it into the chosen feature domain. This transformation process is referred as encoding, whilst decoding corresponds to the reverse action of transforming back into the original domain \cite{coates2011importance_sparsecoding, bengio2013representation}.

Establishing an effective choice of features for the generative model can aid in understanding the nature of the data.
Further, the resulting coefficients after encoding can be used in-place of the raw data for many tasks such as image compression\cite{bryt2008facecompression, horev2012imagecompression, fu2016hyperspectralcompressed}, image denoising \cite{elad2006imagedenoisingsc, achim2001bayesiancauchywaveletus}, image super-resolution \cite{jiang2018ctsrsparse, alejandrof2017cscsuperresolution}; or even image classification \cite{zhang2013lowrankclass, chen2016cscclassification} or anomaly detection \cite{adler2015sparseanomaly} for those requiring some discriminative power in them. % since often the process of the data is carried on within the feature domain due to the easiness of the execution of the task, and the manipulation of the data in general, as opposed to working on the original domain. 

Thus, determining a suitable feature set for the data in question is a crucial task, but how should this be accomplished? Early on, it was common to employ a set of predefined or fixed basis features. Sets such as Wavelets and ones obtained from the Discrete Cosine Transform have been used with success for image denoising and compression respectively. 
%In essence, the data is transformed to the feature domain, in which its  manipulation becomes easier for the task. 
For instance, to denoise images, a common practice is to transform them to the wavelet domain, in which a threshold can be applied to the wavelet coefficients \cite{achim2001bayesiancauchywaveletus}. Reversing the transform to the original domain after this thresholding then results in a cleaner image. However, feature sets such as these are in some sense universal. As such they are not always effective in capturing specific traits of a particular dataset. This can sometimes result in vital particularities of the data being lost. For this reason different approaches enabling the unveiling of new meaningful data specific information have become more prevalent. 

As previously mentioned, the assumptions made will guide the design of the model to utilise for these purposes. Principal component analysis (PCA) \cite{wold1987pca} and independent component analysis (ICA) \cite{bell1995ica} provide the means to determine the underlying components comprising the data of interest. The difference between these two being that ICA makes a further assumption regarding the independence among these components. In either case, the generative model corresponds to a dot product that meets some orthogonality conditions on the (squared) matrix of features. 
There is no strict requirement however, for the feature matrix to be square. For instance, in Autoencoders (AE) \cite{kramer1991autoencoders} the goal is to train a network in an unsupervised fashion such that its weights both encode and decode the data in a lower dimensional space. Alternatively, there is dictionary learning and sparse coding \cite{tosic2011dictionarylearning}, in which it has been suggested that overcomplete sets are capable of describing as well as (if not better than) complete ones as they are able to unveil a bigger number of underlying features \cite{tosic2011dictionarylearning}.  Since the features belong to an overcomplete matrix, the model to solve is underdetermined and an infinity of solutions becomes available. This can be remedied by assuming the data representation is sparse, meaning that only a few elements of the feature matrix take part in the formation of the data and hence most of the elements in the vector of coefficients are set to zero. This is modelled by the addition of a penalty term known to enforce this behaviour. The assumption of sparsity has been motivated by the way in which the V1 cells from the visual cortex work \cite{olshausen1997v1cells}.

With this addition the model not only learns the features to represent the data but also the coefficients describing their contribution. This is commonly achieved by splitting the learning into two tasks, i.e., a step is devoted to learn the elements of the dictionary and the other to learn the elements in the vector of coefficients. There is evidence showing that the latter step can be the most critical for the model to succeed in the representation task \cite{coates2011importance_sparsecoding}. In \cite{coates2011importance_sparsecoding} the power of encoding was demonstrated regardless the choice of learning (or lack of it) for the features. This motivates efforts on the design of novel approaches aiming to learn the coefficients in the encoding stage of the algorithm.

The core contribution of this work is the derivation and effective demonstration of a new regularisation term used during the encoding step of Convolutional Sparse Coding (CSC). This term arises from the assumption that feature map coefficients follow a Cauchy distribution. To make use of this new regularisation we propose the Cauchy proximal operator, which when implemented iteratively follows in the vein of shrinkage algorithms \cite{daubechies2004iterative, blumensath2008iht, ILT} and gives raise to an algorithm, which we refer to as iterative Cauchy thresholding (ICT). Unlike existing previous approaches this algorithm does not perform explicit thresholding, resulting in values approaching 0 but not necessarily locking to it. The power of this new regularisation is shown on a reconstruction task for 2D images, and evaluated against the common choices of soft and hard thresholding algorithms.

The remaining of this manuscript is organised as follows. The backbone and derivation of the proposed algorithm is reviewed in detail in section \ref{sec:proxop} along with related work that inspired our approach. In section \ref{sec:cauchycsc} the algorithm used for the reconstruction task is shown. The experiments conducted are found in section \ref{sec:experiments} along with their results. Lastly, section \ref{sec:conclusions} offers a discussion, conclusion, and future lines of work.

% =======================================================
%     C O N V O L U T I O N A L   S P A R S E   C O D I N G
% =======================================================
%\section{Convolutional Sparse Coding for Representation Learning}
\section{Theoretical Preliminaries}
\label{sec:csc_rl}
In a basic generative model, it is assumed the observations \(\textbf{y} \in \textrm{I\!R}^{M}\) can be estimated from a linear combination of the column vectors (also referred as atoms, codes or features) of the dictionary matrix \(\textbf{A} = [\textbf{a}_1, \textbf{a}_2, ..., \textbf{a}_N] \in  \textrm{I\!R}^{M\times N}\). The contribution of each one of these elements is given by the coefficients in \(\textbf{x} \in \textrm{I\!R}^{N}\) such that there is one coefficient per column in \(\textbf{A}\):

\begin{equation}
    \label{eq:genmodel}
    \begin{aligned}
        \hat{\textbf{y}} &= \textbf{A}\textbf{x} 
    \end{aligned}
\end{equation}

\noindent
Since \(\textbf{y} \approx \hat{\textbf{y}}\), there exists a vector \(\bm{\epsilon}\) such that \(\bm{\epsilon} = \textbf{y} - \hat{\textbf{y}}\) or, equivalently

\begin{equation}
    \label{eq:genmodel_err}
    \begin{aligned}
        \textbf{y}  &= \hat{\textbf{y}} + \bm{\epsilon} \\
    \end{aligned}
\end{equation}

\noindent
with \(\hat{\textbf{y}}, \bm{\epsilon} \in \textrm{I\!R}^M \). For dictionary learning and sparse coding, the dictionary matrix \(\textbf{A}\) is overcomplete \(N > M\). 
In addition, there is a one-to-one spatial correspondence between the features and the data to reconstruct, i.e. the dimension of the features has to be the same as that of the data, which can be impractical for high dimensional signals. This can be alleviated by using patches extracted from the original signal instead, reducing the dimension of the dictionary atoms to the one of these patches. Thus, the observations \(\textbf{y}_i\) corresponds to patches extracted from the original signal of original dimension \(P\) and only \(L\) atoms participate in the generation of the data (\(L \ll N\)). By using patches it becomes necessary to perform pre- and post-processing of the data to extract the patches and then bring them together to reconstruct each sample. For this to work, it is assumed these patches are independent, even if they come from the same sample, which is not necessarily true. There are two main ways to extract patches from the data, one of them is restricting them to not overlap. This, in addition to the independence assumption, is later on reflected in blocking artifacts when the samples are reconstructed. On the other hand, when the patches are overlapped, an average is performed, which also results in a degraded version of the original sample as there is now a smoothing effect present in them. Furthermore, the learned features are often translated versions of other atoms within the set (they are not shift invariant). 

The use of the convolution operator in the generative model helps to address the aforementioned limitations of dictionary learning \cite{zeiler2010deconvolutional, grosse2012shiftcsc}. Thus, it evolves to Convolutional Sparse Coding (CC). Eq. \ref{eq:genmodelconv} describes this model.

\begin{equation}
    \label{eq:genmodelconv}
    \hat{\textbf{y}} = \sum_{k=1}^{K}{\textbf{f}_k * \textbf{z}_k}
\end{equation}

\noindent
where the signal of interest \(\textbf{y} \in \textrm{I\!R} ^ P\) is now modelled as a sum of \(K\) filters \(\textbf{f}_k \in \textrm{I\!R}^M\) convolved with their respective feature map \(\textbf{z}_k \in \textrm{I\!R}^Q\), with \(P = M + Q - 1\) for \(k=1,2,...,K\). Note that \(\textbf{y}\) is the complete original signal. The extension to higher dimensional data is straightforward. Nevertheless, for ease of reading the equations are expressed purely using one dimensional data.

In either of the two mentioned generative models, the learning of the features can be done by minimising the error between the estimated and the observed data. For instance, for CSC:

\begin{equation}
    \label{eq:learngenmodel}
    \begin{aligned}
        \textbf{f}^* &= \arg\underset{\textbf{f}}{\min} \mathcal{L}(\textbf{f}, \textbf{z}) \\
                    &= \arg \underset{\textbf{f}}{\min} ||\textbf{y} - \hat{\textbf{y}}||_{2}^{2} \\
                    &= \arg\underset{\textbf{f}}{\min} ||\textbf{y} - \sum_{k=1}^{K}{\textbf{f}_k * \textbf{z}_k}||_{2}^{2}\\
    \end{aligned}
\end{equation}

\noindent
where \(\textbf{f} = [\textbf{f}_1, \textbf{f}_2, ..., \textbf{f}_K]\) and \(\textbf{z} = [\textbf{z}_1, \textbf{z}_2, ..., \textbf{z}_K]\). In dictionary learning the optimisation would be carried over the matrix \(\textbf{A}\). From now on, the generative model considered in the paper is the CSC.

To seek for sparsity, it suffices to add a regularisation term to the optimisation function as 

\begin{equation}
    \label{eq:learngenmodelreg}
    \begin{aligned}
        \textbf{f}^*, \textbf{z}^*  
            &= \arg\underset{\textbf{f}, \textbf{z}}{\min  } \mathcal{G}(\textbf{f}, \textbf{z}) \\
            &= \arg\underset{\textbf{f}, \textbf{z}}{\min  } \mathcal{L}(\textbf{f}, \textbf{z}) + \lambda \varphi(\textbf{z}) \\
            & \text{s.t. } ||\textbf{f}_k||_2 = 1, k=1,2,...,K
    \end{aligned}
\end{equation}

\noindent
in which it is now required to learn, in addition to the set of features in \(\textbf{f}\), its coefficients in the feature maps \(\textbf{z}_k\). The constraint on the filters prevent it to absorb most of the energy during the learning.

The learning is carried on by iteratively alternating the optimisation of cost function over \(\textbf{f}\) and \(\textbf{z}\). This means that in a first step (z-step), the cost function will be minimised by assuming \(\textbf{f}\) is fixed. The opposite happens during the f-step.

\begin{table*}[!t]
    \renewcommand{\arraystretch}{1.3}
    \caption{Penalty terms that promote sparsity and their proximal operators}
    \label{tab:proximals}
    \centering
    \begin{tabular}{|c|c|c|}
    \hline
    \bfseries Algorithm & \bfseries Penalty term & \bfseries Optimising function \\
    \hline\hline
    \textbf{IHT} 
        & $ |x_i|^0 $
        & $ 
        x_i = 
        \begin{cases}
            x_i, & |x_i| > \lambda \\
            0,   & |x_i| \leq \lambda 
        \end{cases}$ \\
    \hline
    \textbf{IST} 
        &  $ |x_i| $ 
        &  $
        x_i = 
        \begin{cases}
            x_i - \lambda/2 , & x_i > \lambda/2 \\
            x_i + \lambda/2 , & x_i < -\lambda/2 \\
            0,   & otherwise
        \end{cases}$  \\
    \hline
    \textbf{ILT} 
        & \(\lambda \log(\delta + x_i)\) 
        &  
        $
        %\label{eq:proxop_ilt}
            x_i = 
            \begin{cases}
                \frac{1}{2}\left((x - \delta) \pm \sqrt{(x + \delta)^2 -2\lambda)}\right), 
                    & x \geq \sqrt{2\lambda} - \delta \\
                \frac{1}{2}\left((x + \delta) \pm \sqrt{(x - \delta)^2 -2\lambda)}\right), 
                    & x \geq \sqrt{2\lambda} + \delta 
            \end{cases}
        $ 
    \\
    \hline
    \end{tabular}
\end{table*}

% =======================================================
%         I T E R A T I V E   A L G O R I T H M S 
% =======================================================
%\section{Iterative Algorithms for Sparse Coding}
%label{sec:iterative}
Several approaches have emerged aiming to solve Eq. \ref{eq:learngenmodelreg} w.r.t. \(\textbf{f}\) to learn features from the data in addition to Gradient Descent, such as K-SVD \cite{ksvd} and the more image statistic-adapted SparseDT \cite{sparsedt}. Similarly, optimising the cost function w.r.t. \(\textbf{z}\) will result in the learning of the sparse coefficients. Such optimisation depends on the choice of penalty function. If one is to seek for the sparsest solution, then the penalty term chosen is the \(\ell_0\)-norm. Hence, finding the set of coefficients that optimise Eq. \ref{eq:learngenmodelreg} is a combinatorial (NP-hard) problem. Broadly speaking, there are two main approaches to solve said regularisation term: greedy and relaxed algorithms. The first category focuses on solving the \(\ell_0\)-norm whilst the second one considers its relaxed version (the \(\ell_1\)-norm). For the former, Matching Pursuit is one of the most common solvers in which coefficients are chosen one by one in a greedy fashion until a stopping criteria is met. If, on the other hand, one chooses the \(\ell_1\)-norm (LASSO), the function to optimise is now non-smooth convex. 

% The Iterative Shrinkage Thresholding (IST) algorithm \cite{daubechies2004iterative} is a common choice used to solve the \(\ell_1\)-norm. This is defined in Eq. \ref{eq:proxl1}.

%\begin{equation}
%    \label{eq:proxl1}
%    \begin{aligned}
%        \text{prox}_{\lambda\varphi}(x) &= \text{sign}(x) \textrm{max}(0, x - \lambda/2)
%    \end{aligned}
%\end{equation}

In these circumstances, the choice of penalty term is based on the known behaviour (shape) of the function. Thus, as long as one knows the function has a shape that can enforce sparsity, such function can be used for \(\varphi(\cdot)\). Some alternatives are the non-convex \(\ell_p\)-norm (with \(p \leq 1\)) or the (also non-convex) log regulariser among others. A comprehensive review of these sparsity-enforcing functions can be found in \cite{wen2018surveynonconvex} and references therein. 

Iterative algorithms have come along with an associated sparsity enforcing penalty term. IST is often involved when the function to optimise makes use of the \(\ell_1\)-norm, IHT for the \(\ell_0\)-norm and recently the use of the iterative log thresholding (ILT) \cite{ILT} has been proposed to optimise the log regulariser. Table \ref{tab:proximals} summarises the equations involved in these algorithms.

IST and IHT can also be derived via surrogate functions in which one seeks to separate the terms involved in the cost function. Regardless of the chosen algorithm to use, these thresholding operators are applied in an element-wise fashion. These three approaches suppress any value below some threshold but it is only IST and ILT that update values higher than said threshold. In the case of IST this has a direct impact on the results as they often exhibit blurring.

% It is of particular interest, however, the algorithm Iterative Log Thresholding (ILT) proposed in \cite{ILT}, which shares a few similarities with the one presented in this paper. The log regulariser is defined in Eq. \ref{eq:ilt_pen}.

%\begin{equation}
%    \label{eq:ilt_pen}
%    \begin{aligned}
%       \varphi(x) &= \lambda \log(\delta + x)
%    \end{aligned}
%\end{equation}

%\noindent
%, which, similarly to the \(\ell_1\)-norm, can be solved by using its proximal operator (Eq. \ref{eq:proxop_ilt}).

%\begin{equation}
%    \label{eq:proxop_ilt}
%    z = 
%    \begin{cases}
%        \frac{1}{2}\left((x - \delta) \pm \sqrt{(x + \delta)^2 -2\lambda)}\right), 
%            & x \geq \sqrt{2\lambda} - \delta \\
%        \frac{1}{2}\left((x + \delta) \pm \sqrt{(x - \delta)^2 -2\lambda)}\right), 
%            & x \geq \sqrt{2\lambda} + \delta 
%    \end{cases}
%\end{equation}

% =======================================================
%     C A U C H Y   P R O X I M A L   O P E R A T O R 
% =======================================================
\section{Iterative Cauchy Thresholding}
\label{sec:proxop}
The Cauchy assumption in the field of image processing is not new as it has previously been used to model the noise corrupting the images of interest \cite{mei2018cauchynoiseremoval, sciacchitano2015variational}. Nonetheless, in this work it is not the noise but the coefficient values involved in the generative models the ones that are assumed to follow this distribution. In fact, the assumption of a Cauchy prior for the model has been done with success in the past % in the wavelet domain, in which it is known the data does not follow the Gaussian distribution
\cite{wan2011segmentation_bivariate_cauchy, achim2005image, bhuiyan2007spatially, ranjani2010dual, gao2013directionlet}.

The encoding step depends on the regularisation term in the optimisation model. This term could fall into the non-smooth convex functions, such as the \(\ell_1\)-norm; non-smooth non-convex, such as the log regulariser or the \(\ell_0\)-pseudo norm; or smooth non-convex penalty terms, such as the one explored in this paper. This function is derived from a statistical assumption on the coefficients, serving as prior in a maximum a posteriori (MAP) approach. The resultant learning algorithm corresponds to a function which, despite the non-convexity of its regularisation term, is guaranteed to convergence under a certain condition. Specifically, it is the Cauchy distribution the one assumed to drive the learning framework. The use of this prior enables the learning of the coefficients by iteratively applying its proximal operator on the coefficients, achieving shrinkage around a non-explicit threshold that emerges naturally from the equations involved in this process. In addition, the parameters shaping the distribution of the coefficients can be estimated from the observations, facilitating the use of this method.

\subsection{The Cauchy distribution}
\label{subsec:cauchy}
\begin{figure}
    \centering
    \includegraphics[scale=0.6]{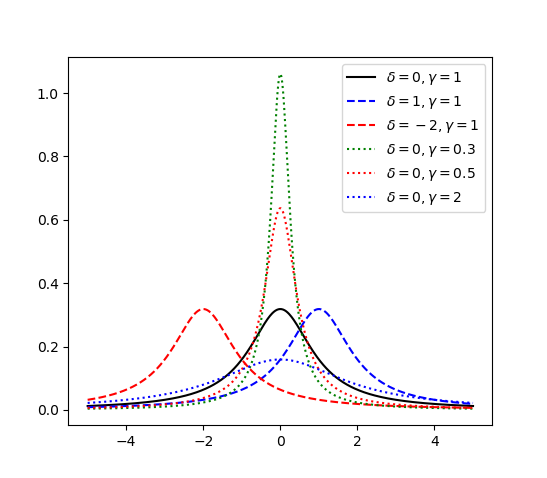}
    \caption{Cauchy PDF with different values for the parameters \(\delta\) and \(\gamma\)} \label{fig:cauchypdfs}
\end{figure}

The Cauchy distribution belongs to the family of the Symmetric \(\alpha\)-Stable (S\(\alpha\)S ) distribution. Its location and dispersion are described by the parameters \(\delta\) and \(\gamma\) respectively, and its p.d.f. is defined by 

\begin{equation}
    \label{eq:cauchypdf}
    p(x) = \frac{\gamma}{\pi(\gamma^2 + (x-\delta)^2)}
\end{equation}
whilst Figure \ref{fig:cauchypdfs} illustrates their role on  the distribution.

For the aim of the proposed work (enforcing sparsity), it is required that \(\delta = 0\), which in turn simplifies the expression to work with. On the other hand, the parameters of the distribution can be estimated from the data itself using maximum likelihood estimation as %done in Eq. \ref{eq:estimgamma}.
\begin{equation}
    \label{eq:estimgamma}
    \begin{aligned}
		\gamma &= \arg\underset{\gamma}{\min  } -\sum_{t=1}^{T} \log(p(x) + \epsilon) \\
	\end{aligned}
\end{equation}
where \(\epsilon\) is a very small value.

%\subsection{Proximal splitting algorithms}
%\label{subsec:proxalg}
The overall cost function to be optimised is composed of two functions, as illustrated in Eq. \ref{eq:learngenmodelreg}. The data fidelity term \(\mathcal{L}(\cdot, \cdot)\) being convex with \(\varphi(\cdot)\) possibly non-smooth and/or non-convex. In proximal splitting, the functions that present challenges during conventional optimisation techniques are projected into a convex set via their proximal operators. Note that the number of functions involved in the optimisation can be \(\geq 2\). The proximal operator of a given function  \(\varphi(\cdot)\) can be obtained by solving:
\begin{equation}
    \label{eq:proxop}
    \begin{aligned}
		\text{prox}_{\lambda\varphi}(x) &= \arg\underset{\textbf{z}}{\min}
		& & (z-x)^2 + \lambda \varphi (z) \\
	\end{aligned}
\end{equation}

%Hence, it is required to solve Eq. \ref{eq:proxop} for \(z\).

\subsection{The Cauchy Proximal Operator}
\label{subsec:ict}
Following a MAP approach, the penalty term on the coefficients in \(\textbf{z}\) is then defined as \(\varphi(\cdot) = -\log(p(\cdot))\), with \(p(\cdot)\) as defined in Eq. \ref{eq:cauchypdf} and setting \(\delta = 0\). This penalty term is applied individually on every element of \(\textbf{z}\). Thus, plugging in this into Eq. \ref{eq:proxop} it is possible to derive the Cauchy proximal operator as%, requiring only to minimise Eq. \ref{eq:cauchy_proxop} over \(z\):
\begin{equation}
    \label{eq:cauchy_proxop}
    \begin{split}
        \text{prox}_{\lambda \varphi}(x) &= \arg\min_{\textbf{z}} \quad(z - x)^2 + \lambda \varphi(z) \\
         &= \arg\min_{\textbf{z}} \quad (z - x)^2 - \lambda\log\left(\frac{\gamma}{\pi \left(\gamma^2 + z^2 \right) }\right)
    \end{split}
\end{equation}

Taking the derivative to find the stationary points:
\begin{equation*}
    \frac{d}{dz}\left((z - x)^2 - \lambda\log\left(\frac{\gamma}{\pi \left(\gamma^2 + z^2 \right) }\right)\right)=0
\end{equation*}

\begin{equation*}
    \frac{d}{dz}\left((z - x)^2\right) - \lambda\frac{d}{dz}\left(\log\left(\frac{\gamma}{\pi \left(\gamma^2 + z^2 \right) }\right)\right)=0
\end{equation*}

\begin{equation*}
    2(z - x) - \lambda\frac{d}{dz}\left(-\log\left(\gamma^2 + z^2 \right)\right)=0
\end{equation*}

\begin{equation*}
    2(z - x) + \lambda\frac{d}{dz}\left(\log\left(\gamma^2 + z^2 \right)\right)=0
\end{equation*}

\begin{equation*}
    2(z - x) + \lambda\frac{1}{\left(\gamma^2 + z^2 \right)} \frac{d}{dz}\left(\gamma^2 + z^2 \right)=0
\end{equation*}

\begin{equation*}
    2(z - x) + \lambda\frac{2z}{\left(\gamma^2 + z^2 \right)}=0
\end{equation*}

\begin{equation*}
    (\gamma^2 + z^2)(z-x) + \lambda z = 0
\end{equation*}

Lastly, rearranging the terms, %for better display 
we get

\begin{equation}
    \label{eq:ict_min}
    z^3 - x z^2 + (\gamma^2+ \lambda) z - \gamma^2 x = 0
\end{equation}

Using the Cardano's method to find the roots of the previous cubic equation with \(a=1\), \(b=-x\), \(c= \gamma^2 + \lambda\) and \(d = -\gamma^2 x\), one finally gets to:

\begin{equation}
    \label{eq:cauchy_shrinkage}
    z = \frac{x}{3} + t %+ s
\end{equation}

\noindent
where

\begin{equation*}
    \begin{split}
        t &= \sqrt[3]{-\frac{q}{2}+\sqrt[2]{\Delta}} + \sqrt[3]{-\frac{q}{2}-\sqrt[2]{\Delta}}\\
        % s &=  \\
        \Delta &= \frac{q^2}{4}+\frac{p^3}{27} \\
        p &= \lambda + \gamma^2-\frac{x^2}{3} \\
        q &= -\frac{2}{27}x^3 + \frac{1}{3}\left(\lambda-2\gamma^2\right)x
    \end{split}
\end{equation*}

\begin{figure}[!t]
    \centering
    \subfloat[]{\includegraphics[width=1.7in]{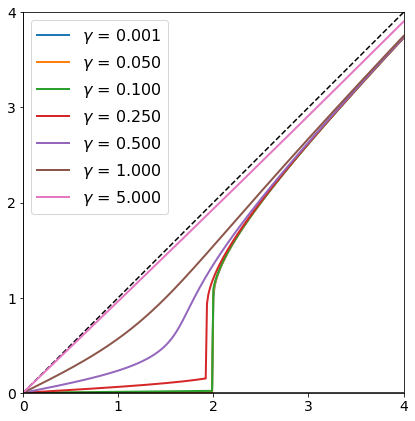}
    \label{fig:diff_gammas_ict}}
    %\hfil
    \subfloat[]{\includegraphics[width=1.7in]{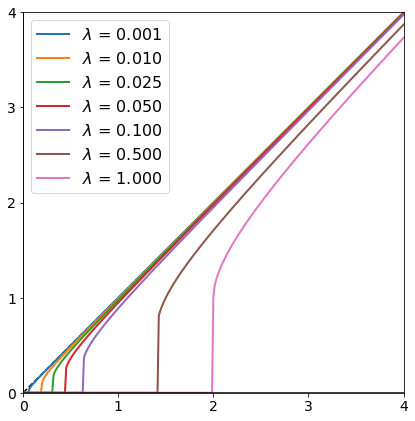}
    \label{fig:diff_lmbdas_ict}}
    \caption{Behaviour of ICT for varying (a) \(\gamma\)'s and (b) \(\lambda\)'s.}
    \label{fig:ict_behaviour}
\end{figure}

%This will provide only one real root when \(\Delta > 0\), and three real roots otherwise. In the latter case the root with the largest absolute value is chosen. 

The Cauchy proximal operator, thus, requires to choose values for the parameters \(\gamma\) and \(\lambda\), for which it becomes ideal to understand their function in the operator. By fixing \(\gamma\) to a specific value and then vary \(\lambda\) and vice-versa it is possible to gain an intuition of their roles. In fact, it is found that the value of \(\gamma\) shapes the thresholding function and smaller values contribute to a more aggressive shrinkage near the threshold, whilst \(\lambda\) shifts the threshold location.  Fig. \ref{fig:diff_gammas_ict} and \ref{fig:diff_lmbdas_ict} illustrate this behaviour. 

In fact, when \(\gamma \rightarrow 0\) the threshold \(\rightarrow 2\lambda\) and the shape of the function approximates the ILT. On the other hand, when \(\lambda \rightarrow 0\), the roots of Eq. \ref{eq:ict_min} \(\rightarrow \gamma i, -\gamma i\), and \(x\), which would keep the values unchanged, i.e., not shrinkage would be performed. Nonetheless, in this work we do not treat \(\gamma\) as a tunable parameter. Instead, this value is estimated from the data following the approach mentioned in Section \ref{subsec:cauchy}.

One could compare the Cauchy and log penalty terms (third row in Table \ref{tab:proximals}) since both are shaped by the logarithm function and some parameter, \(\delta\) for ILT and \(\gamma\) for ICT. The corresponding proximal operators are considerably different. A major difference between ICT and the rest of the algorithms presented in this manuscript so far is the lack of an explicit threshold. The coefficients are still shrunk according to the proximal operator in an iterative manner, reaching values closer to zero but not necessarily locking on it.

\begin{figure}
    \centering
    \includegraphics[width=0.4\textwidth]{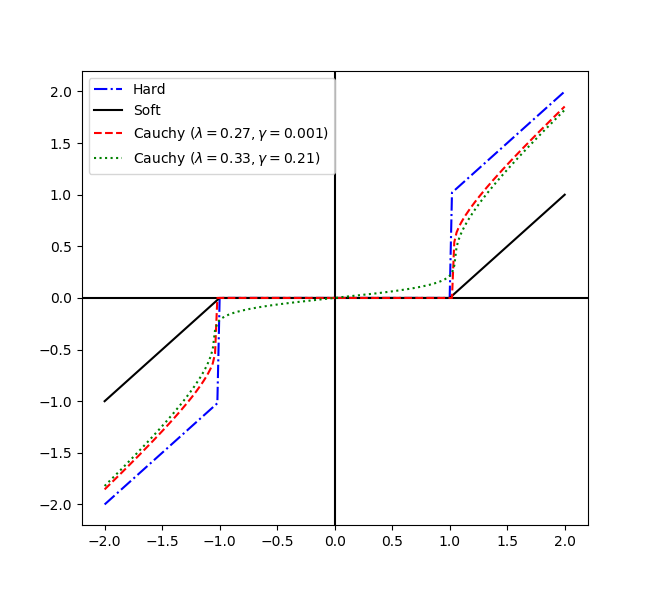}
    \caption{Behaviour of the different thresholding algorithms and, in the case of Cauchy, using different parameters.}
    \label{fig:diff_thresholding}
\end{figure}

%The Cauchy proximal operator has been previously derived and used successfully on various tasks \cite{loza2010bivariate_cauchy, wan2011segmentation_bivariate_cauchy}. Our approach differs in the starting point of such derivation. In the aforementioned works the proximal operator emerges from a Bayesian approach, whilst this one considers the forward backward splitting method.
%The proximal operator derived from the Cauchy distribution has been previously used in various signal processing tasks. 

%Nonetheless, the achievements of these works under the Cauchy assumption should not be neglected and, moreover, they encourage and motivate even further the approach proposed in this work.
%In \cite{loza2010bivariate_cauchy} and \cite{wan2011segmentation_bivariate_cauchy}, the bivariate Cauchy distribution was used to define a shrinkage algorithm able to process the wavelet coefficients for the tasks of denoising and segmenting images. Even though the penalty term has been also obtained from a Bayesian approach, the shrinkage thresholding algorithm has been derived from a proximal operator perspective in such a way it is implemented in a proximal splitting approach, which can also be seen as forward-backward splitting algorithm. This proximal operator has also been recently explored in \cite{oktay2020cauchy} % [Cite the other papers written by Oktay and myself?]. 

The penalty term derived from the Cauchy distribution is a smooth non-convex function, which makes the optimisation of \(\mathcal{G}(\cdot,\cdot)\) challenging . However, the cost function defined in Eq. \ref{eq:learngenmodelreg}, as a whole, can be guaranteed to converge to a global minimum, if the following condition is met \cite{oktay2020cauchy}:

\begin{equation}
    \label{eq:cauchy_condition}
    \begin{aligned}
		\lambda \leq 8\gamma^2 \\
	\end{aligned}
\end{equation}

Specifically, the condition given in Eq. \ref{eq:cauchy_condition} ensures that the cost function in the Cauchy proximal operator (\ref{eq:cauchy_proxop}) converges. %It has also been shown that 
This condition guarantees convergence in the scenario in which the proximal operator needs to be applied in an iterative manner for inverse problems \cite{oktay2020cauchy} %\cite{oktay2020cauchy, karakucs2020solving,karakucs2020covid19}. 
It is this iterative process the one that gives rise to  our proposed ICT algorithm, whose pseudocode is presented in Algorithm \ref{alg:ict}. Note that an additional parameter \(\eta\) is present as it accounts for the learning rate, thus, the original equation contains \(\eta\lambda\), and since \(\lambda = 1\) the algorithm has \(\eta\) only, which also affects the convergence condition to \(\eta \leq 8\gamma^2 \) instead.

\begin{algorithm}[htbp]
    \begin{algorithmic}
     %   \SetAlgoLined
     %   \KwResult{Sparse coefficients \(\textbf{x}\)}
        \STATE Initialise \(\textbf{x}\) with 0's\;
        \STATE Set \(\eta\), \(\gamma\) and \(\lambda\)\;
        \STATE Choose stopping criterion. In this work this corresponds to a max number of iterations\;
        \WHILE {Stopping criteria has not been met}
            \STATE Compute \(\textbf{z} \gets \textbf{z} - \eta\nabla_\textbf{z}\mathcal{L}(\textbf{f}, \textbf{z})\)\;
            \STATE Shrink every element in \(\textbf{z}\) using Eq. (\ref{eq:cauchy_shrinkage}).
        \ENDWHILE
    \end{algorithmic}
    \caption{Iterative Cauchy Thresholding}
    \label{alg:ict}
\end{algorithm}

This condition is easily applied when the generative model corresponds to CSC. As shown by \cite{oktay2020cauchy}, this arises from the condition being derived by taking the second derivative where the generative model is no longer involved and the resultant expression is dependant only on the hyper-parameters.

%In the following section, this result is extended to the specific case of CSC, in which the argument of the shrinkage operator contains the convolutional generative model, since the Hessian is not dependent on it. 

% =======================================================
%  C A U C H Y   C O N V O L U T I O N A L   S P A R S E   C O D I N G % =======================================================
\vfill\null
\section{Cauchy Convolutional Sparse Coding}
\label{sec:cauchycsc}

\begin{figure*}
    \centering
    \includegraphics[width=0.71\textwidth]{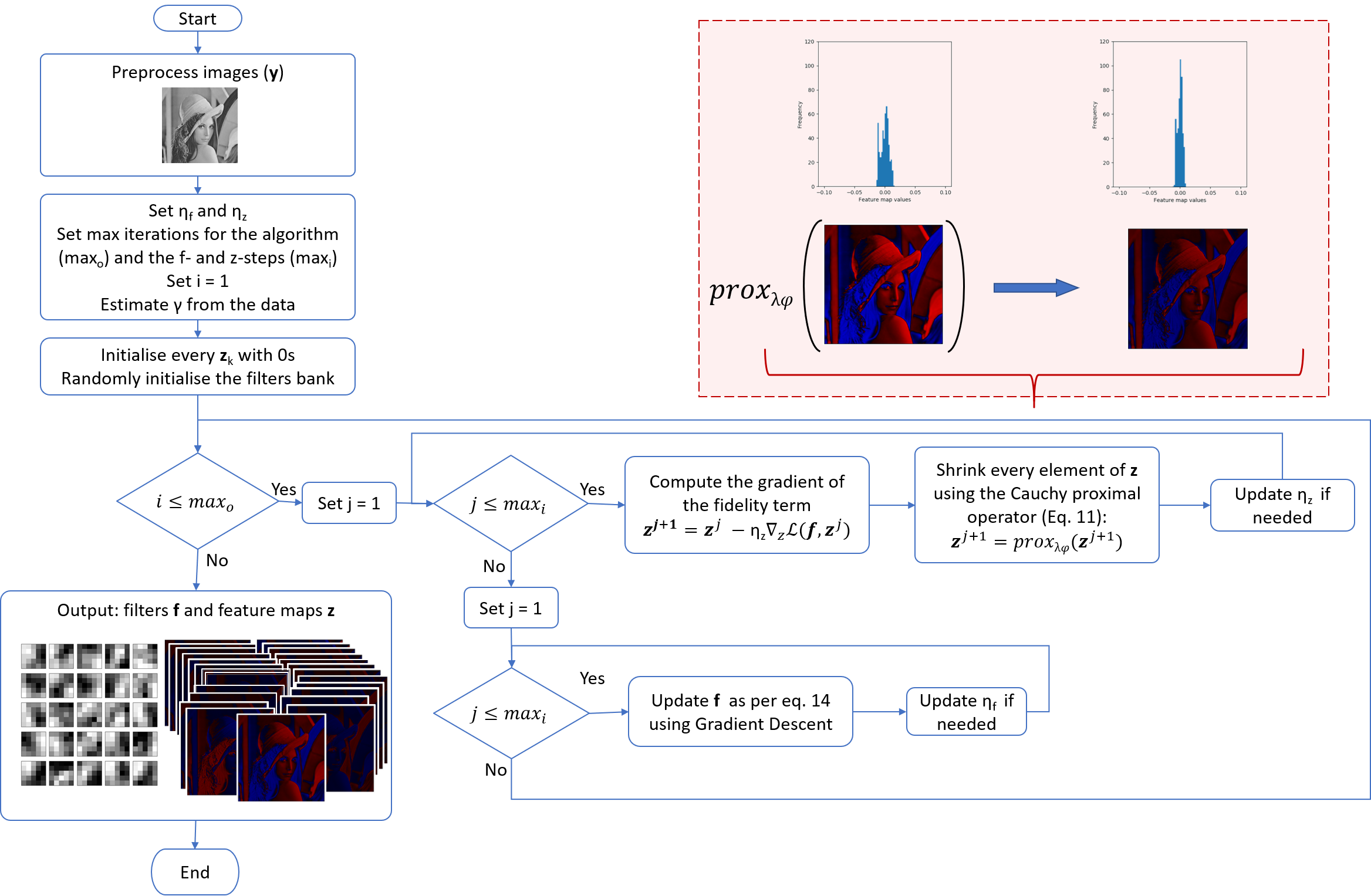}
    \caption{Block diagram of the Cauchy Convolutional Sparse Coding algorithm. After a few iterations the coefficients within the feature maps get closer to zero.}
    \label{fig:ccsc_diagram}
\end{figure*}

In this section, our new CSC algorithm for representation learning is introduced. It is based on the use of the Cauchy proximal operator through an iterative process in order to encode the data for the \(z\)-step.  
The cost function is derived via MAP. The prior knowledge employed and which then translates into the penalty function corresponds to the assumed statistical distributions of the coefficients \cite{grosse2012shiftcsc}.

By using the Cauchy distribution in the generative model it is now required to perform the \(z\)-step via the Cauchy proximal operator. Our goal is thus to solve Eq. (\ref{eq:learngenmodelreg}) for the feature maps using the Cauchy penalty function. Specifically, the cost function is now:

\begin{equation}
    \label{eq:cauchy_cost}
    \mathcal{G}(\textbf{f}, \textbf{z}) = ||\textbf{y}-\hat{\textbf{y}}||_2^2 - \lambda \sum_{k=1}^{K}\sum_{q=1}^{Q}\log\left(\frac{\gamma}{\pi(\gamma^2 + z_{k,q}^2)}\right)
\end{equation}

\noindent
with \(\hat{\textbf{y}}\) as defined in Eq. \ref{eq:genmodelconv}. 
Thus, the full algorithm aims to learn the set of filters \(\textbf{f}\) and the feature maps \(\textbf{z}\) associated to the data from a dataset of size \(T\). Note that extending the cost function defined in \ref{eq:learngenmodelreg} to learn from more than one sample (i.e. dataset size \(>\) 1) is straightforward and hence not detailed in here.

As it is common in similar algorithms, the proposed approach works by alternating between the learning of the features and the learning of the coefficients, until a convergence criterion is met. This can consist in reducing the reconstruction error below some predefined value or in a maximum number of iterations to be reached. Gradient descent is used as learning approach for the features (f-step) in conjunction with a chosen thresholding algorithm for the coefficients (z-step). In both learning steps the learning rate adapts so that overshooting over the local minima is prevented. This is achieved by halving the learning rate for the current step following an observed increase in value subsequent to an update. 

\begin{algorithm}[ht!]
    \begin{algorithmic}
     %   \SetAlgoLined
     %   \KwResult{Sparse coefficients \(\textbf{x}\)}
        \STATE Initialise \(\textbf{z}_k\) with 0's, \(k = 1, 2, .., K\)\;
        \STATE Initialise randomly \(\textbf{f}_k\), \(k = 1, 2, .., K\) \;
        \STATE Estimate \(\gamma\) from the data using Eq. \ref{eq:estimgamma}
        \STATE Choose stopping criteria\;
        \WHILE {Overall stopping criteria has not been met}
            \STATE Set \(\textbf{z}^{old} \gets \textbf{z}\) \;
            \WHILE {Stopping criteria for \(z\)-step has not been met}
                \STATE Set \(C_O \gets \mathcal{G}(\textbf{f}, \textbf{z})\)\;
                \STATE For every \(\textbf{z}_k\) compute:\\ \(\textbf{z}_k \gets \textbf{z}_k - \eta_z\nabla_{\textbf{z}_k}\mathcal{L}(\textbf{f}, \textbf{z})\)\;
                \STATE Shrink \(\textbf{z}_k\) using Eq. (\ref{eq:cauchy_shrinkage}).
                \STATE Set \(C_N \gets \mathcal{G}(\textbf{f}, \textbf{z}) \)
                \IF{ \(C_N > C_O\)}
                    \STATE Set \(\textbf{z} \gets \textbf{z}^{old}\) \;
                    \STATE Set \(\eta_z = \eta_z / 2\)
                \ELSE
                    \STATE Set \(\textbf{z}^{old} \gets \textbf{z}\) \;
                \ENDIF
            \ENDWHILE
            \WHILE {Stopping criteria for \(f\)-step has not been met}
                \STATE Set \(C_O \gets \mathcal{G}(\textbf{f}, \textbf{z})\)\;
                \STATE For every \(\textbf{f}_k\) compute GD on the filers: \\ \(\textbf{f}_k \gets \textbf{f}_k - \eta_f\nabla_{\textbf{f}_k}\mathcal{L}(\textbf{f}, \textbf{z})\)\;
                \IF{ \(C_N > C_O\)}
                    \STATE Set \(\textbf{f} \gets \textbf{f}^{old}\) \;
                    \STATE Set \(\eta_f = \eta_f / 2\)
                \ELSE
                    \STATE Set \(\textbf{f}^{old} \gets \textbf{f}\) \;
                \ENDIF
            \ENDWHILE
        \ENDWHILE
    \end{algorithmic}
    \caption{Cauchy Convolutional Sparse Coding}
    \label{alg:ccsc}
\end{algorithm}

The \(f\)-step is solved by minimising Eq. \ref{eq:cauchy_cost} over \(\textbf{f}\), which can be written compactly as 

\begin{equation}
    \label{eq:learnoverfilt}
    \begin{aligned}
        \textbf{f}^* &= \arg\underset{\textbf{f}}{\min}\quad ||\textbf{y} - \sum_{k=1}^{K}{\textbf{f}_k * \textbf{z}_k}||_{2}^{2}
    \end{aligned}
\end{equation}

This requires taking the gradient over \(\textbf{f}\) and choosing a step size \(\eta_f\) for updating the features iteratively. This guarantees convergence since it involves the optimisation of the \(\ell_2\)-norm, which is a smooth convex function.
%The overshoot of the fixed point is avoided by updating the learning rate as previously explained.

%\subsection{Learning features from data}
%\label{subsec:learnfeatures}

\begin{figure*}[!t]
    \centering
    \subfloat[]{\includegraphics[width=1.66in]{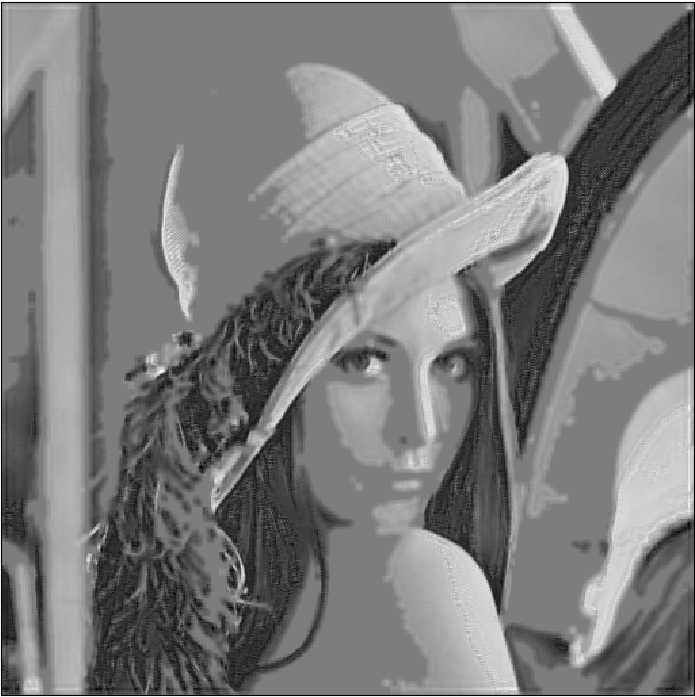}}
    %\label{fig:hard_recons}}
    \hfil
    \subfloat[]{\includegraphics[width=1.66in]{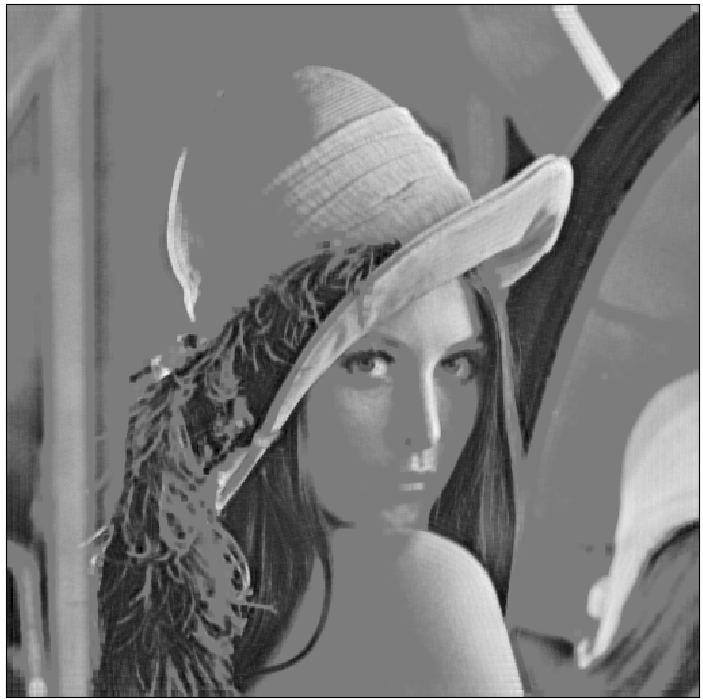}}
    %\label{fig:laplace_recons}}
    \hfil
    \subfloat[]{\includegraphics[width=1.66in]{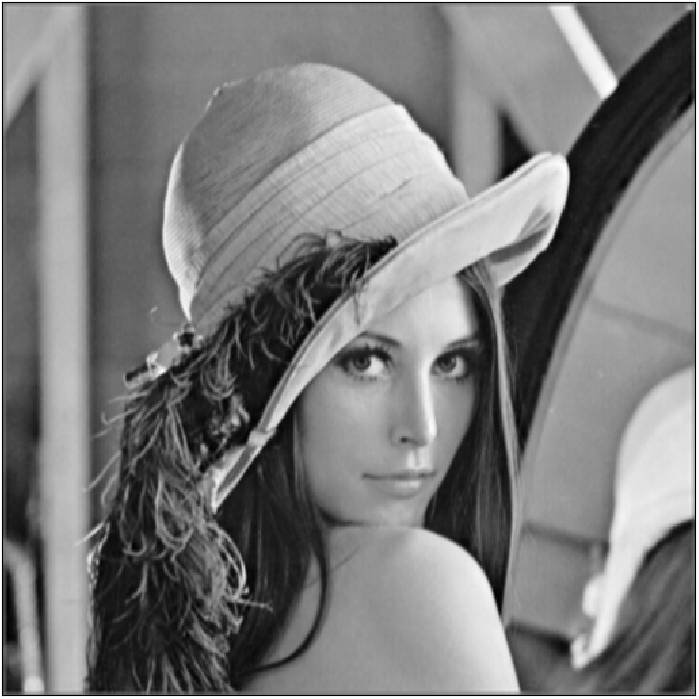}}
    %\label{fig:cauchy_recons}}
    \\
    \subfloat[]{\includegraphics[width=1.66in]{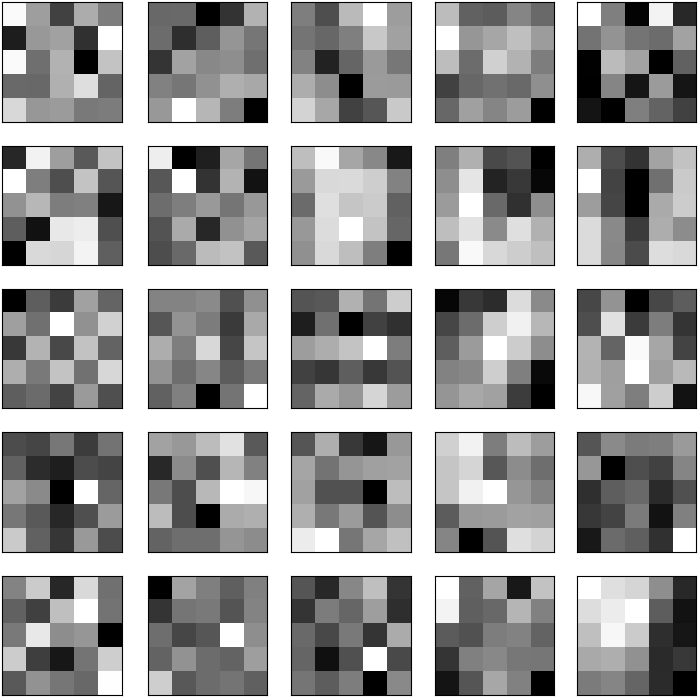}}
    %\label{fig:hard_filt}}
    \hfil
    \subfloat[]{\includegraphics[width=1.66in]{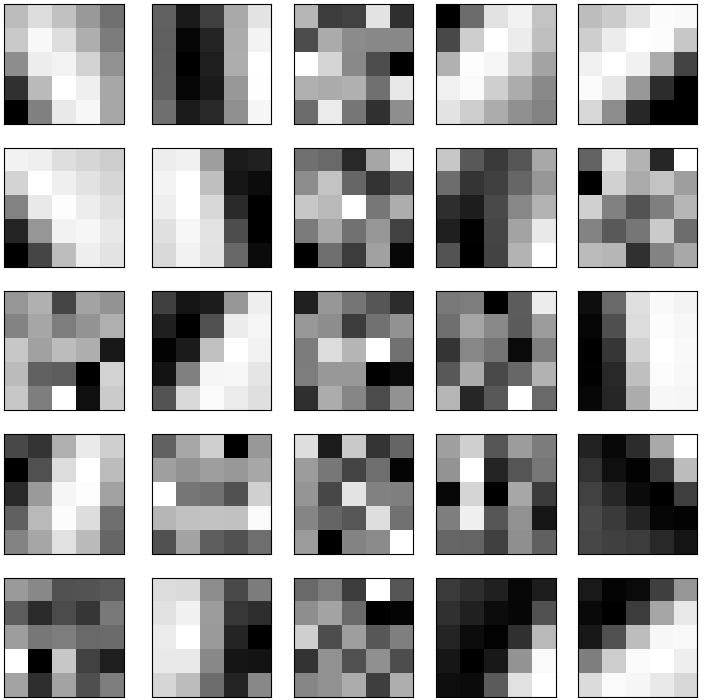}}
    %\label{fig:laplace_filt}}
    \hfil
    \subfloat[]{\includegraphics[width=1.66in]{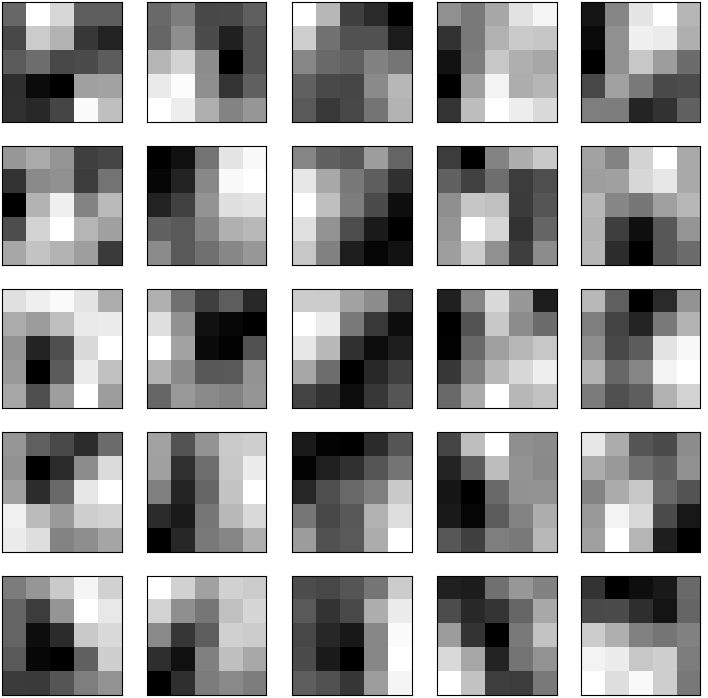}}
    %\label{fig:cauchy_filt}}
    \\
    \subfloat[]{\includegraphics[width=1.66in]{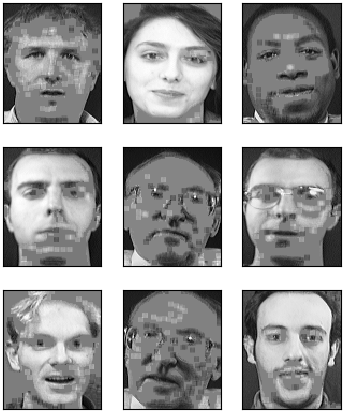}}
    %\label{fig:hard_recons}}
    \hfil
    \subfloat[]{\includegraphics[width=1.66in]{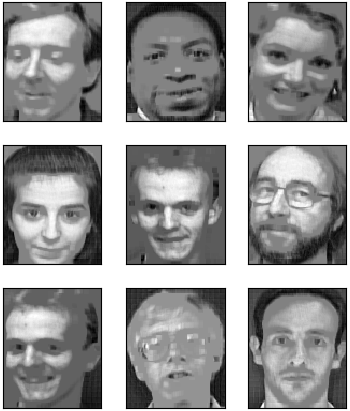}}
    %\label{fig:laplace_recons}}
    \hfil
    \subfloat[]{\includegraphics[width=1.66in]{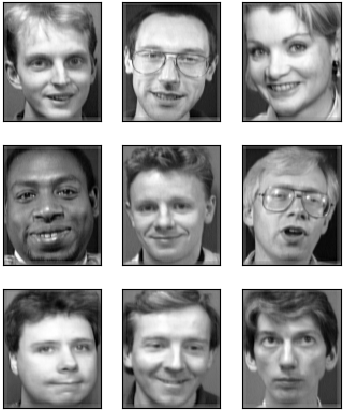}}
    %\label{fig:cauchy_recons}}
    \\
    \subfloat[]{\includegraphics[width=1.66in]{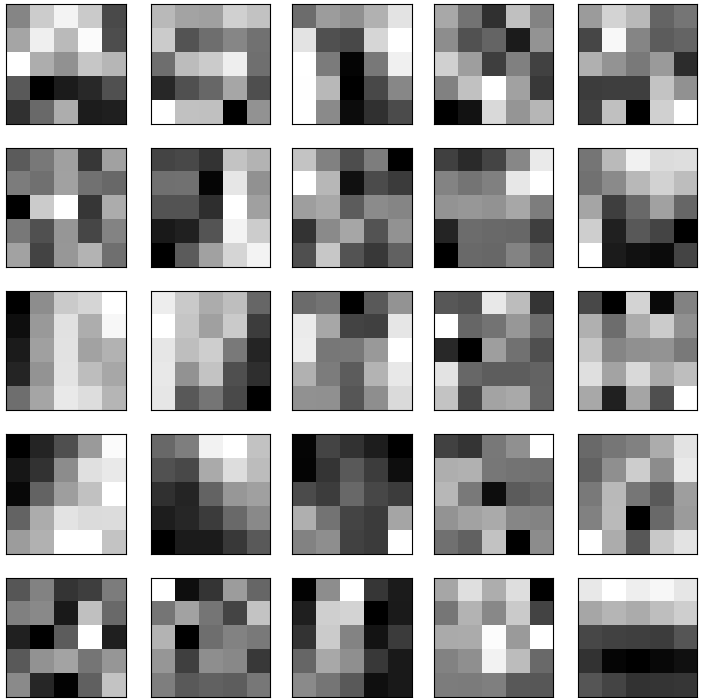}}
    %\label{fig:hard_filt}}
    \hfil
    \subfloat[]{\includegraphics[width=1.66in]{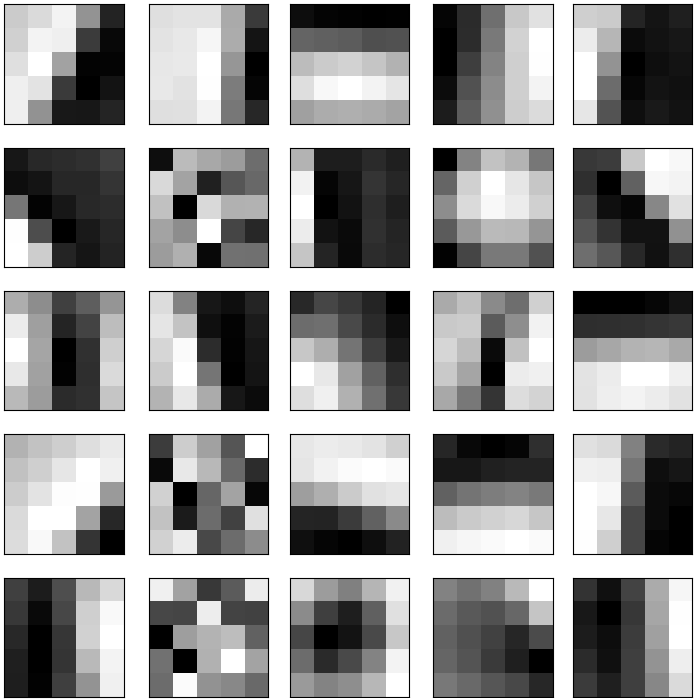}}
    %\label{fig:laplace_filt}}
    \hfil
    \subfloat[]{\includegraphics[width=1.66in]{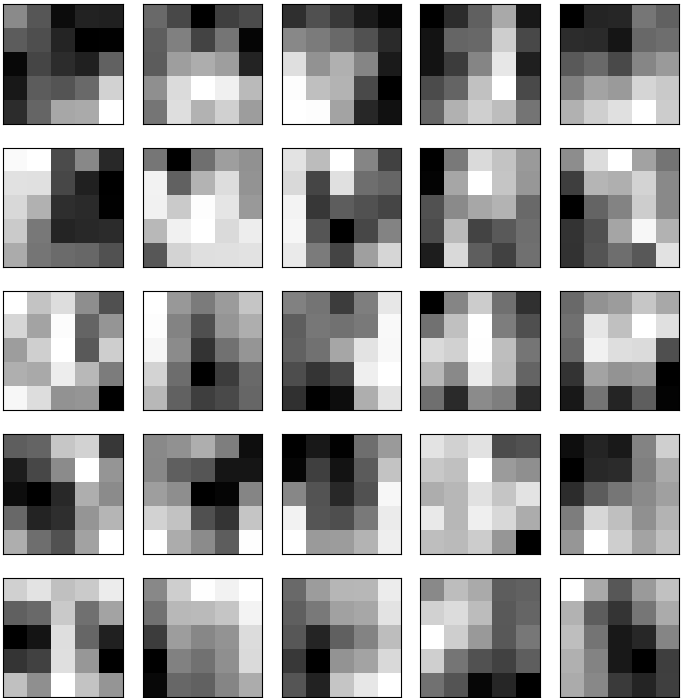}}
    %\label{fig:cauchy_filt}}
    \\
    \caption{Reconstructions of image Lena (top row) using the algorithms (a) IHT, (b) IST and (c) ICT and the filters learned using (a) IHT, (b) IST and (c) ICT.}
    \label{fig:filters_lena}
\end{figure*}

\begin{figure*}[!t]
    \centering
    \subfloat[]{\includegraphics[width=1.66in]{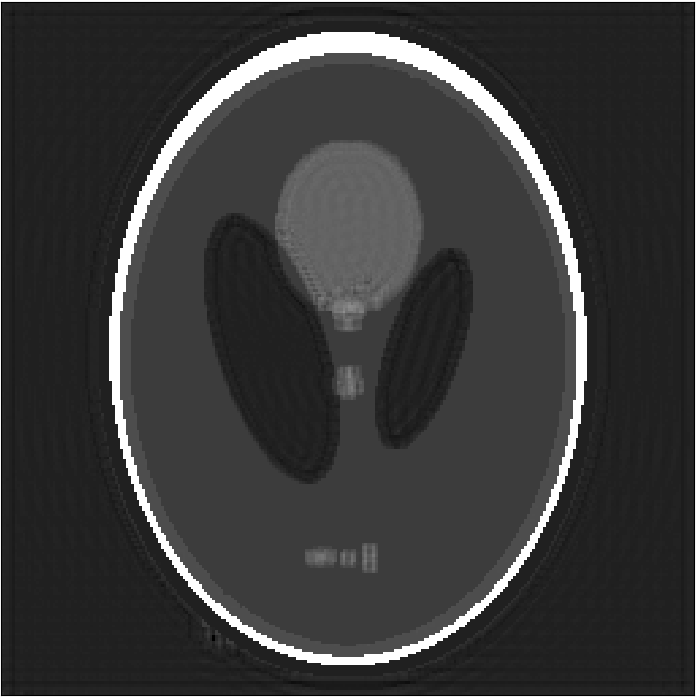}}
    %\label{fig:hard_recons}}
    \hfil
    \subfloat[]{\includegraphics[width=1.66in]{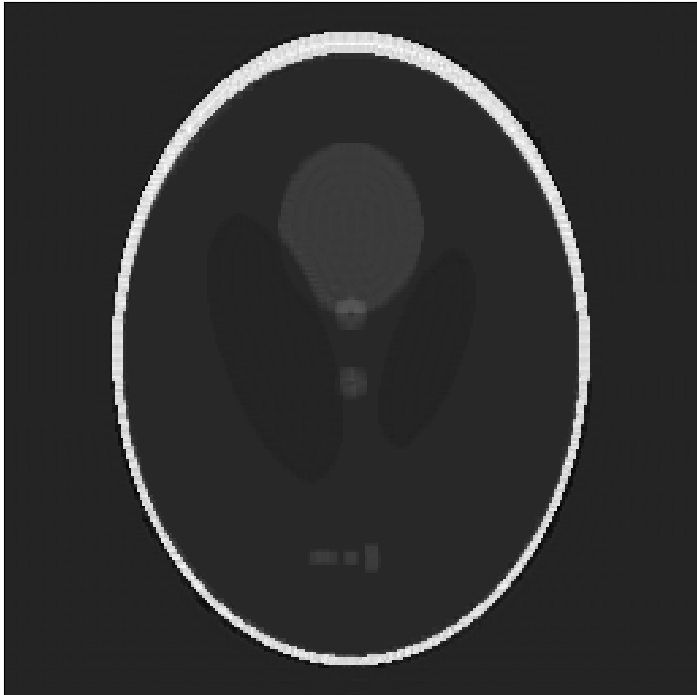}}
    %\label{fig:laplace_recons}}
    \hfil
    \subfloat[]{\includegraphics[width=1.66in]{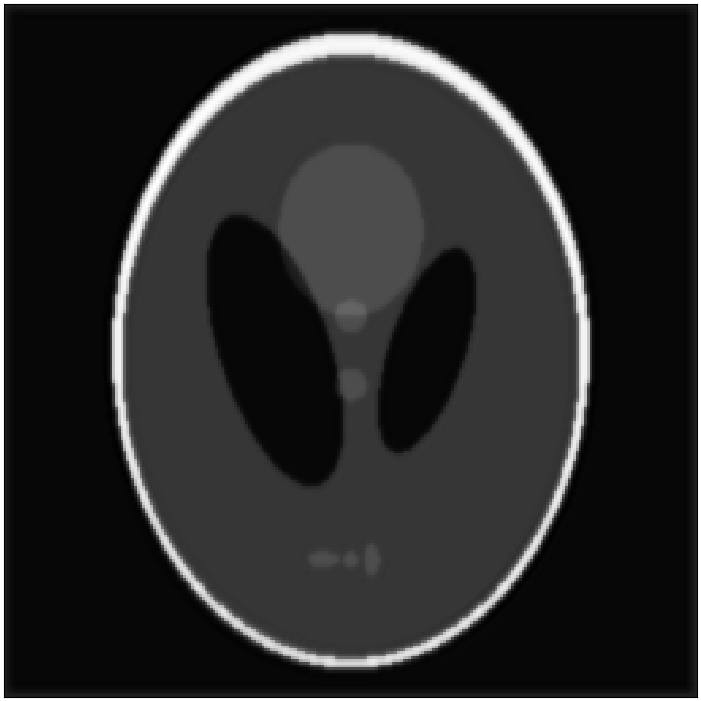}}
    %\label{fig:cauchy_recons}}
    \\
    \subfloat[]{\includegraphics[width=1.66in]{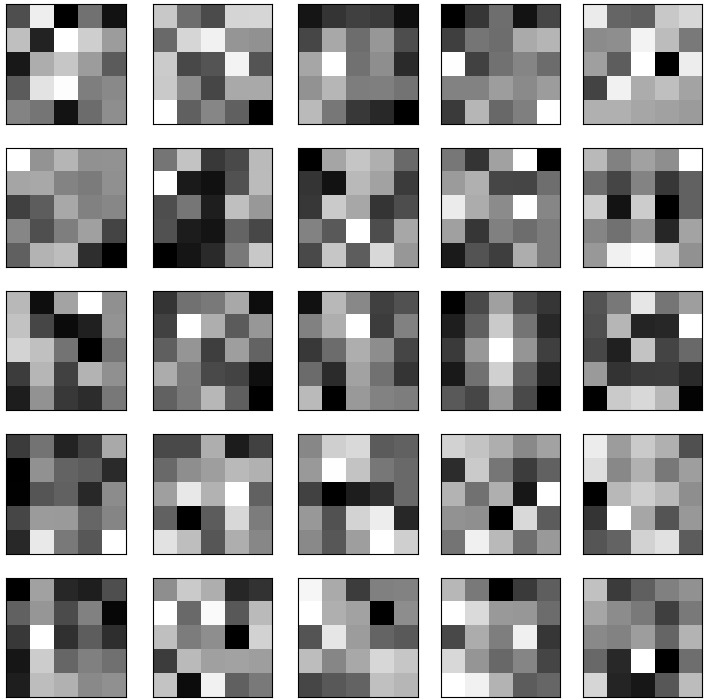}}
    %\label{fig:hard_filt}}
    \hfil
    \subfloat[]{\includegraphics[width=1.66in]{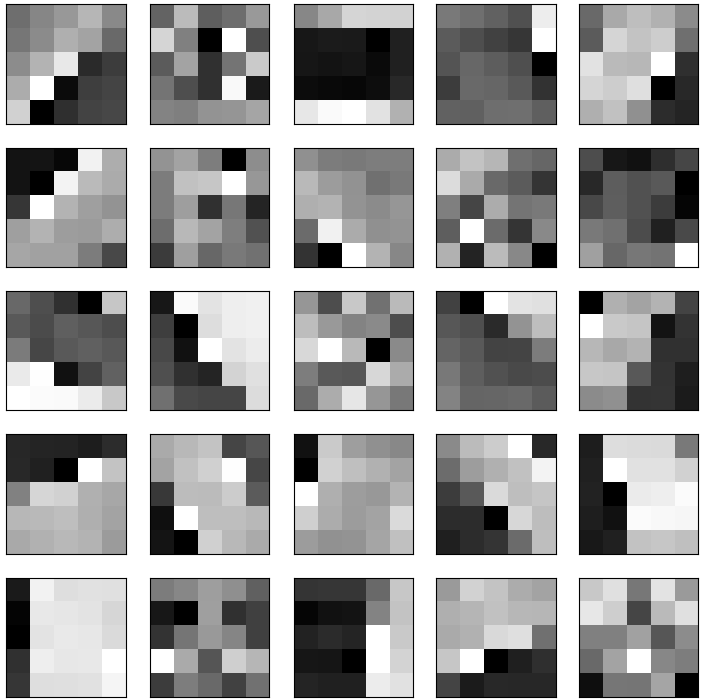}}
    %\label{fig:laplace_filt}}
    \hfil
    \subfloat[]{\includegraphics[width=1.66in]{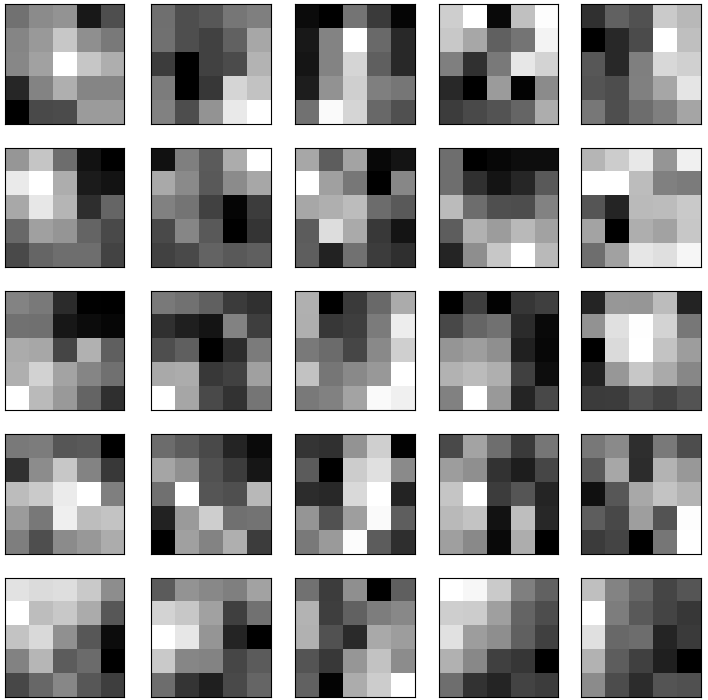}}
    %\label{fig:cauchy_filt}}
    \\
    \subfloat[]{\includegraphics[width=1.66in]{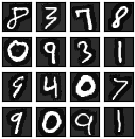}}
    %\label{fig:hard_recons}}
    \hfil
    \subfloat[]{\includegraphics[width=1.66in]{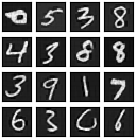}}
    %\label{fig:laplace_recons}}
    \hfil
    \subfloat[]{\includegraphics[width=1.66in]{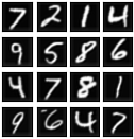}}
    %\label{fig:cauchy_recons}}
    \\
    \subfloat[]{\includegraphics[width=1.66in]{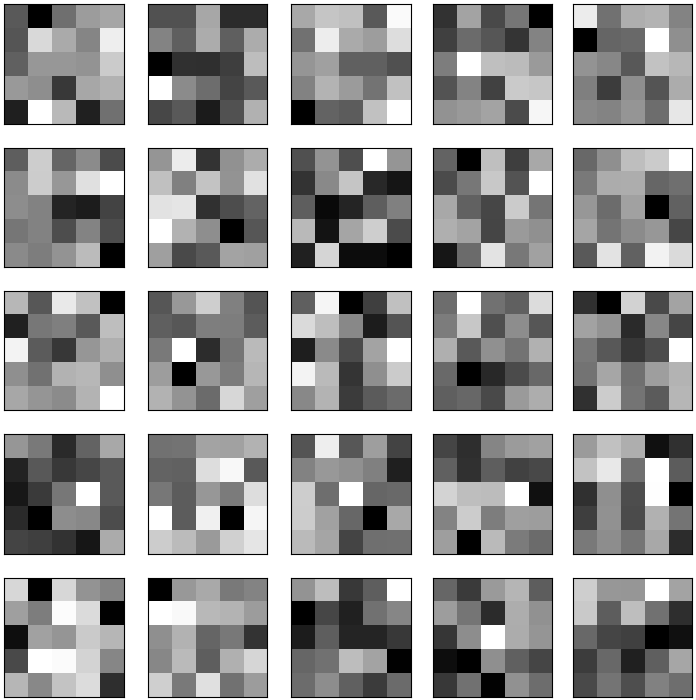}}
    %\label{fig:hard_filt}}
    \hfil
    \subfloat[]{\includegraphics[width=1.66in]{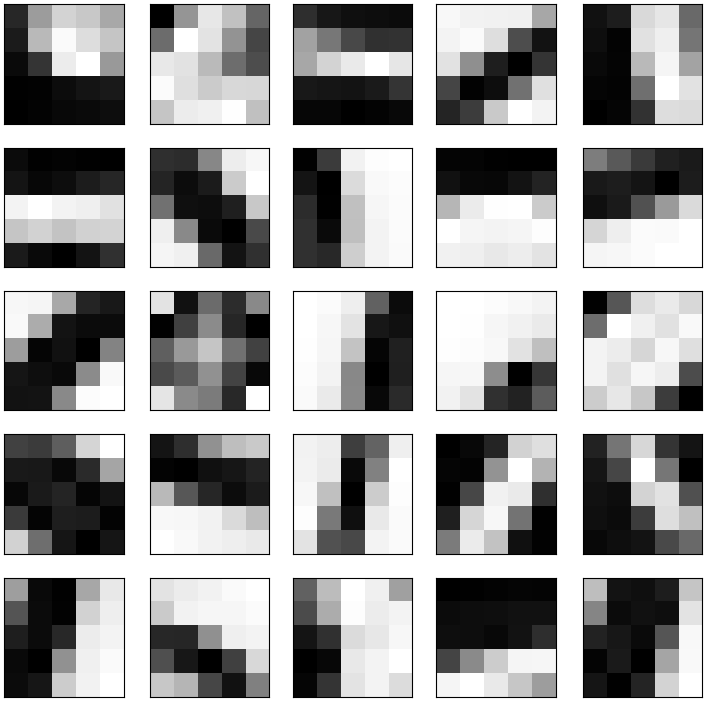}}
    %\label{fig:laplace_filt}}
    \hfil
    \subfloat[]{\includegraphics[width=1.66in]{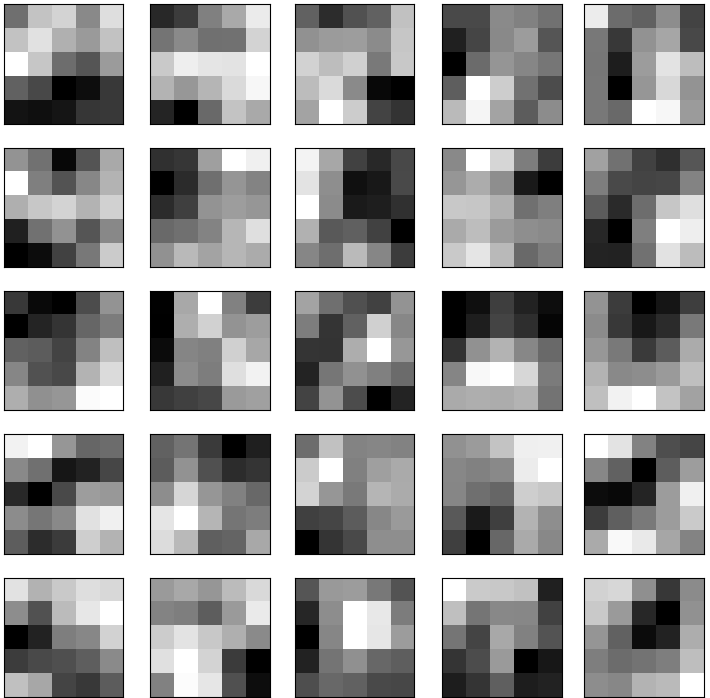}}
    %\label{fig:cauchy_filt}}
    \\
    \caption{Reconstructions of image Shepp-Logan Phantom (top row) using the algorithms (a) IHT, (b) IST and (c) ICT and the filters learned using (a) IHT, (b) IST and (c) ICT.}
    \label{fig:filters_phantom}
\end{figure*}

%\subsection{Learning the coefficients}
%\label{subsec:learncoeff}
%The regularisation term involved in the optimisation enables the selection of the features and their contribution in the reconstruction model, hence its relevance in the algorithm. 
%Iterative thresholding algorithms are commonly used for this step. In this work, 
\clearpage
The key to implementing the Cauchy-CSC (CCSC) method consists in using ICT in the encoding phase of the algorithm. This is achieved by solving:

\begin{equation}
    \label{eq:learnovercoeff}
    \begin{aligned}
        \textbf{z}^* &= \arg\underset{\textbf{z}}{\min}\quad ||\textbf{y} - \sum_{k=1}^{K}{\textbf{f}_k * \textbf{z}_k}||_{2}^{2} \\
        &- \lambda \sum_{k=1}^{K}\sum_{q=1}^{Q}\log\left(\frac{\gamma}{\pi(\gamma^2 + z_{k,q}^2)}\right)
    \end{aligned}
\end{equation}

In addition to the regularisation parameter \(\lambda\), one requires also to choose a learning rate \(\eta_z\). Similarly to what has been done for the \(f\)-step, \(\eta_z\) is updated whenever the cost function increases as result of the previous coefficient  update. The pseudocode of the whole approach is presented in Algorithm \ref{alg:ccsc} and its diagram is depicted in figure \ref{fig:ccsc_diagram}.

% =======================================================
%               E X P E R I M E N T S
% =======================================================
\section{Simulation Results}
\label{sec:experiments}
In order to assess the performance of CSC when used in conjunction with ICT, IHT and IST we conducted a number of experiments. In particular, we focused on the reconstruction of 2D images in order to quantify the results of the said algorithms. The data employed were classical images such as Lena and the Shepp-Logan phantom, as well as the MNIST and AT\&T faces\footnote{\url{ https://git-disl.github.io/GTDLBench/datasets/att_face_dataset/}} datasets. Before applying the representation learning algorithm, independently of the regulariser used, the data have been pre-processed to make them zero-mean. There is no pre-processing done to enforce the dataset to have unit variance since this could affect the estimation of the \(\gamma\) parameter required by the ICT algorithm. Since MNIST and the faces dataset are considerably large, a sample composed of \(T=500\) and \(T=30\) random images therein were used in the respective experiments. 

The complete approach was performed 100 times for each dataset using different random initialisation for the filters. For the MNIST and AT\&T datasets a random set of samples was also chosen at the beginning of each of their experiments. The maximum number of iterations was fixed to 100 per experiment. 

Note that the hyperparameter \(\lambda\) incorporates the learning rate for IHT and IST, whereas for ICT  it was set to 1 in order to leave it as close to the original cost function derived from MAP as possible. Hence, only the estimation of \(\gamma\) is required. %Nevertheless, the choice of the learning rate for both steps is however required. 

For ICT, the learning rate needs to meet the condition in Eq. \ref{eq:cauchy_condition}. The additional tunable parameters employed were $K=25$ and a filter size of 5$\times$5 for all the experiments.

% =======================================================
%               R E S U L T S
% =======================================================
%\section{Results}
%\label{sec:results}
\vfill\null
For an initial qualitative assessment, Figures \ref{fig:filters_lena} and \ref{fig:filters_phantom} show the filters learned for the different datasets, along with the reconstructed images. We show samples from the experiments with the highest PSNR for each algorithm. By visually inspecting Figures \ref{fig:filters_lena} and \ref{fig:filters_phantom}, it can be seen that ICT can learn more meaningful filters since they seem to present less random patterns, in contrast to IST and IHT, in which some of these bases failed to be updated. In fact, we noticed that the initialisation of the filters plays an important role in their learning as sometimes there seem to be no learning at all for IHT as the filters have a noisy appearance. This is in spite of their relatively good reconstruction performance with high PSNR values achieved and this confirms the dependence of reconstruction performance on the encoding step \cite{coates2011importance_sparsecoding}.

\begin{figure*}[!t]
    \centering
    \subfloat[]{\includegraphics[width=3in]{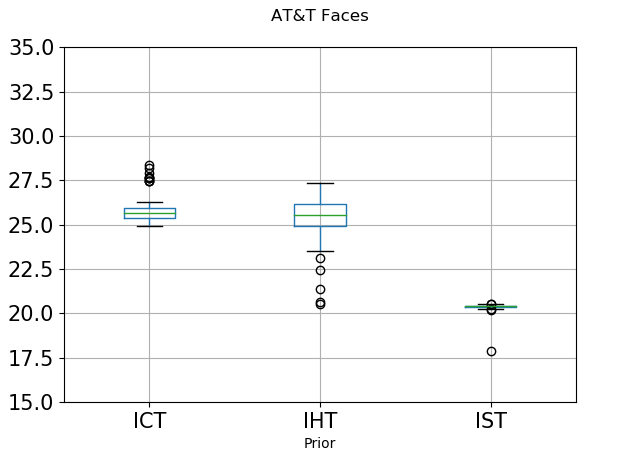}
    \label{fig:att_boxplot}}
    \hfil
    \subfloat[]{\includegraphics[width=3in]{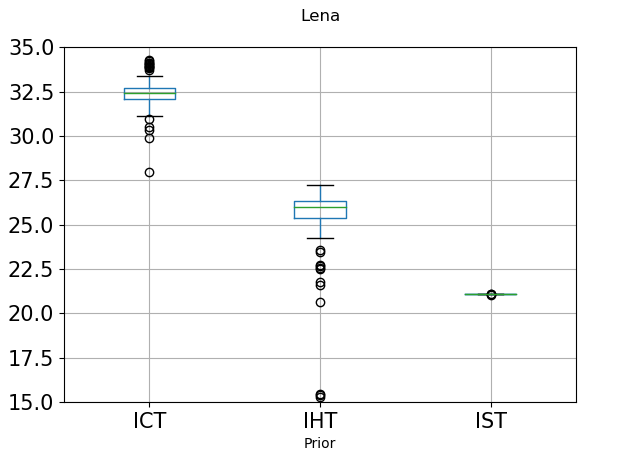}
    \label{fig:lena_boxplot}}
    \hfil
    \\
    \subfloat[]{\includegraphics[width=3in]{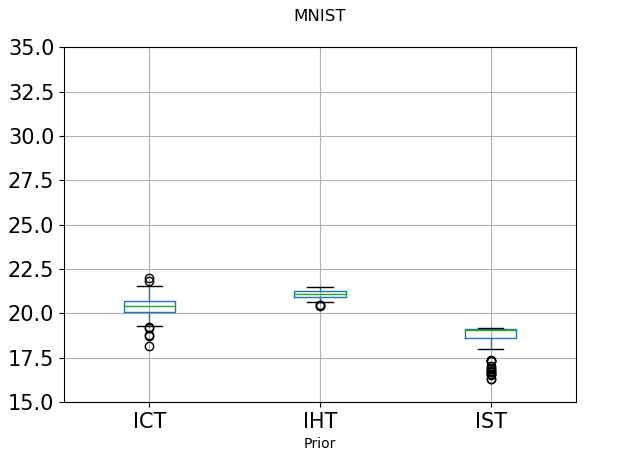}
    \label{fig:mnist_boxplot}}
    \hfil
    \subfloat[]{\includegraphics[width=3in]{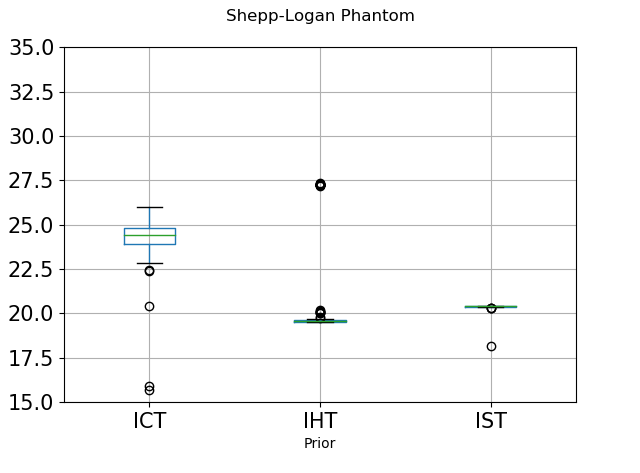}
    \label{fig:phantom_boxplot}}
    \hfil
    \caption{Boxplots of PSNR for CSC using different algorithms for z-step.}
    \label{fig:boxplots_psnr}
\end{figure*}

The performance of the three representation learning approaches is also assessed through quantitative analysis. The PSNR values for the reconstruction of each sample was computed and their average values are reported in Table \ref{tab:results_psnr} and Fig. \ref{fig:boxplots_psnr} showing their respective boxplots. Table \ref{tab:results_sparsity} reports the average proportion of non-zero elements in the learned feature maps. In both table \ref{tab:results_psnr} and table \ref{tab:results_sparsity}, the best performance for each dataset is shown in bold.

\begin{figure*}[!t]
    \centering
    \subfloat[]{\includegraphics[width=3.5in]{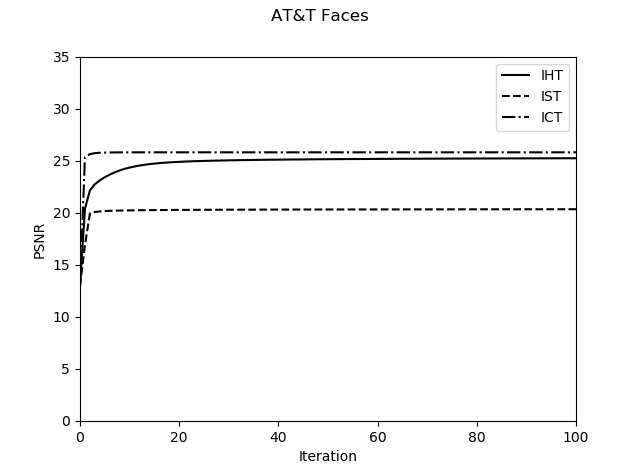}
    \label{fig:att_iter_avg}}
    \hfil
    \subfloat[]{\includegraphics[width=3.5in]{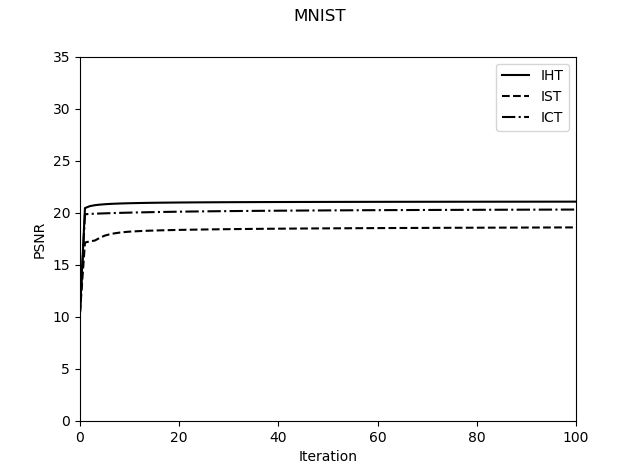}
    \label{fig:lena_iter_avg}}
    \hfil
    \\
    \subfloat[]{\includegraphics[width=3.5in]{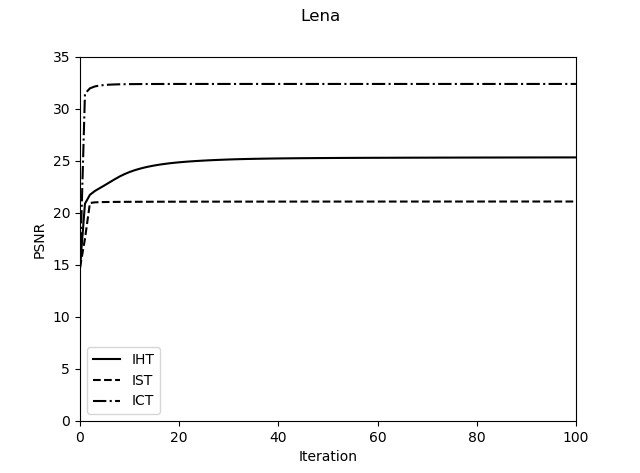}
    \label{fig:mnist_iter_avg}}
    \hfil
    \subfloat[]{\includegraphics[width=3.5in]{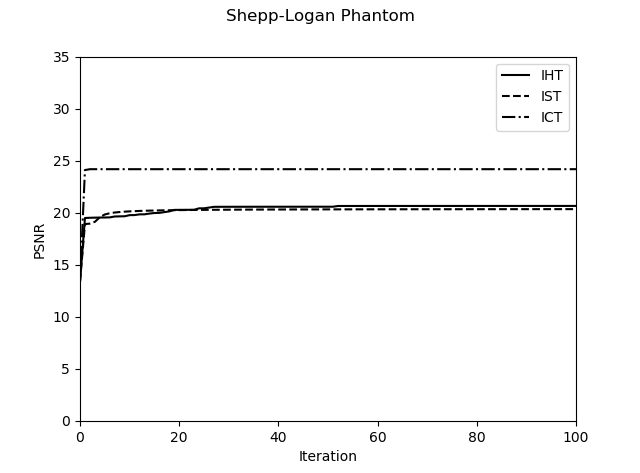}
    \label{fig:phantom_iter_avg}}
    \hfil
    \caption{Learning iterations vs PSNR for CSC using different algorithms for z-step.}
    \label{fig:iter_learn_psnr}
\end{figure*}

Lastly, in Fig. \ref{fig:iter_learn_psnr}, a plot of the learning performance as function of average PSNR as the iterations progress is provided.

\begin{table}[!t]
    \renewcommand{\arraystretch}{1.3}
    \caption{PSNR results of CSC using different penalty terms}
    \label{tab:results_psnr}
    \centering
    \begin{tabular}{|c|c|c|c|}
    \hline
    \bfseries Dataset & \bfseries ICT & \bfseries IHT & \bfseries IST \\
    \hline\hline
    \textbf{MNIST} & 20.36 & \textbf{21.09} & 18.57 \\
    \hline
    \textbf{AT\&T} Faces & \textbf{25.82} & 25.31 & 20.37 \\
    \hline
    \textbf{Phantom} & \textbf{24.20} & 20.67 & 20.37 \\
    \hline
    \textbf{Lena} & \textbf{32.39} & 25.35 & 21.09 \\
    \hline
    \end{tabular}
\end{table}

\begin{table}[!t]
    \renewcommand{\arraystretch}{1.3}
    \caption{Proportion of zero coefficients learned via CSC using different penalty terms}
    \label{tab:results_sparsity}
    \centering
    \begin{tabular}{|c|c|c|c|}
    \hline
    \bfseries Dataset & \bfseries ICT & \bfseries IHT & \bfseries IST \\
    \hline\hline
    \textbf{MNIST} & 99.99 & 5.65 & \textbf{1.16} \\
    \hline
    \textbf{AT\&T} Faces & 99.99 & 2.91 & \textbf{1.65} \\
    \hline
    \textbf{Phantom} & 100 & 2.60 & \textbf{1.37} \\
    \hline
    \textbf{Lena} & 99.99 & \textbf{1.57} & 1.60 \\
    \hline
    \end{tabular}
\end{table}

%\section{Discussion}
%\label{sec:discussion}

From Fig. \ref{fig:iter_learn_psnr} we can see that ICT and IST reach the plateau corresponding to the highest PSNR early in the learning process, with IHT reaching its own maximum a few iterations later. It is ICT, however, the one that achieves the highest PSNR and requires the least iterations. Both IHT and IST requires tuning of a number of parameters for optimal performance, whereas for ICT the parameter $\gamma$ is estimated directly from the data. In fact, the use of the iterative Cauchy algorithm requires the choice of only two values, the learning rate and the scale parameter. As noted in section \ref{subsec:ict}, \(\gamma\) can be estimated from the original data whilst \(\eta_z\) needs to obey the condition in Eq. \ref{eq:cauchy_condition}.

From Table \ref{tab:results_psnr} we can see that ICT provides the best PSNR performances in three out of four cases, which is consistent with the visual evaluation. IST, on the other hand, is the one with the worst reconstruction performances, although it leads to the sparsest representations and the learning of seemingly sharper features in comparison to ICT. IST presents more consistency in regard of the PSNR results obtained as the inter-quartiles range is shorter than the other two algorithms, as seen in Fig. \ref{fig:boxplots_psnr}. %A possible cause of this is the fact that IST shrinks every coefficient on every iteration, whereas IHT keeps higher values untouched. In the Case of ICT, higher values are barely changed and those values close to its natural threshold are shurnk in a more abrupt manner. It can be said, thus, the performance of ICT is found in between the ones of IHT and IST.

There is an increase of the average PSNR as the images increase in size for the three thresholding algorithms considered, with ICT exhibiting the highest of such jumps. Indeed, ICT performed better as the size of the images increased, having very similar performance to IHT for the small size MNIST (28$\times$28) dataset as opposed to the case of the the Shepp-Logan phantom (256$\times$256) and Lena (512$\times$512) images. Furthermore, despite the high PSNR values obtained with IHT, some degradation in the reconstructed images surrounding the edges and the lose of details is apparent. In the case of IST, the images exhibit a smoothing effect regardless of their dimension. Lastly, the reconstructed images produced by ICT also present some artifacts near the edges, which become more apparent in smaller image sizes.

\begin{figure}[t!]
    \centering
   \includegraphics[width=3.25in]{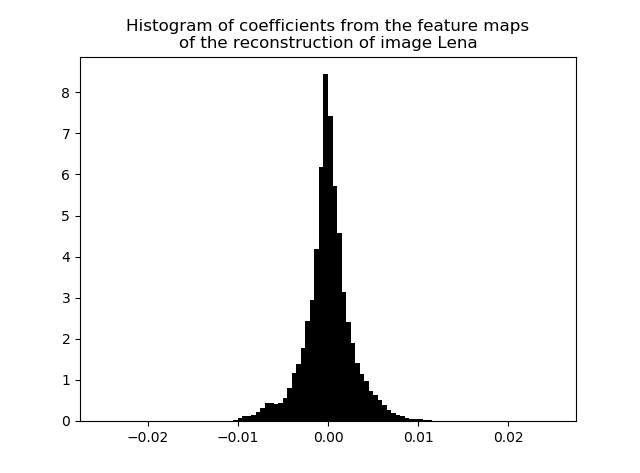}
    \caption{Histogram of coefficient values from the 25 feature maps involved in the reconstruction of the image Lena using ICT. Y axis scale factor: \(10^5\).}
    \label{fig:hist_cauchy}
\end{figure}

With respect to Table \ref{tab:results_sparsity}, it is evident that ICT is the approach that offers the least sparse solutions. Having a closer look to the histograms of coefficients (Fig. \ref{fig:hist_cauchy}) it can be observed that most coefficients are in a very close vicinity of 0, which might explain the ability to learn most of the features most of the times whilst reducing their noisy appearances.

% =======================================================
%               C O N C L U S I O N S
% =======================================================
\section{Conclusions and Future Work}
\label{sec:conclusions}
In this work a new convolutional sparse coding framework based on a Cauchy model assumption is proposed. This approach enables the learning of filters and their respective feature maps by using said distribution as prior for the coefficients in the latter ones, which results in a new cost function. The Cauchy proximal operator was derived and used to optimise it and this requires a preliminary step before the learning process, which involves the estimation of the corresponding scale parameter. The performance was evaluated on four different datasets and compared against the reconstruction performance achieved using hard and soft thresholding. Even though CCSC does not achieve the same degree of sparsity as IST and IHT, the filters learned are seemingly better for the reconstruction task based on their higher PSNR values achieved. Current work focuses on investigating the discriminative power of the proposed representation in classification problems.

\ifCLASSOPTIONcaptionsoff
  \newpage
\fi



%\bibliographystyle{IEEEtran}
\bibliographystyle{IEEEtran}
\bibliography{references}
\end{document}